\documentclass{proc}

\usepackage{defs}
\usepackage{natbib}
\newcommand{\kr}{Kr\"{u}ger 60}

\usepackage{hyperref}
\hypersetup{
    bookmarks=true,                                  
    pdftitle={Eclipsing binary systems as tests of low-mass stellar evolution theory},          
    pdfauthor={Gregory A. Feiden},                   
    colorlinks=true,                                 
    linkcolor=black,                                  
    citecolor=black,                                  
    urlcolor=blue                                    
}

\editors{Miloslav Zejda}
\publisher{}
\conference{Living Together: Planets, Host Stars, and Binaries}
\conferencedate{2014}

\shortauthors{Gregory A. Feiden}
\shorttitle{Eclipsing binaries as tests of low--mass stellar evolution theory}

\title{Eclipsing binary systems as tests of low--mass stellar evolution theory}

\author{Gregory~A.~Feiden$^1$}
\affiliation{1}{Uppsala University, Uppsala, Sweden, gregory.feiden@physics.uu.se}

\abs{
Stellar fundamental properties---masses, radii, effective temperatures---can be
extracted from observations of eclipsing binary systems with remarkable precision,
often better than 2\%. Such precise measurements afford us the opportunity to 
confront the validity of basic predictions of stellar evolution theory, such 
as the mass--radius relationship. A brief historical overview of confrontations 
between stellar models and data from eclipsing binaries is given, highlighting key
results and physical insight that have led directly to our present understanding.
The current paradigm that standard stellar evolution theory is insufficient to 
describe the most basic relation, that of a star's mass to its radius, along the 
main sequence is then described. Departures of theoretical expectations
from empirical data, however, provide a rich opportunity to explore various 
physical solutions, improving our understanding of important stellar astrophysical 
processes.}

\begin{document}

\maketitle

\section{Introduction}
Eclipsing binary (EB) systems 
are some of the best tools for testing stellar evolution theory. Advancements in stellar 
evolution theory motivated by confrontations between stellar model predictions and observations 
of EBs are second, perhaps, only to studies of globular cluster color-magnitude diagrams. 
The power of EB observations to further stellar evolution
theory is derived from observer's ability to extract accurate stellar masses and radii of 
the individual components. Being that mass is the primary input parameter in stellar models
and radius is a primary output quantity, EBs permit direct comparison with model predictions
at the most fundamental level. Furthermore, the presence of two, presumably coeval, stars 
in a binary system forces stellar models to predict the properties of both stars at the 
same age, effectively making each binary system a miniature star cluster.

This review will attempt to characterize advances in low-mass stellar evolution theory
that were and are motivated by investigations of EBs. Naturally, the definition of a low-mass
star is strongly dependent on context, so for reference, I will refer to low-mass stars as
stars with masses below 0.8~\msun. Stars in the range of 0.8 to 1.2 \msun\ will be referred
to as ``solar-like'' stars. These definitions are largely set by the characteristics of 
stellar evolution models in the given mass regimes. The lower boundary of 0.8 \msun\ is defined
by the growing importance of non-ideal contributions to the gas equation of state and a need
for detailed radiative transfer models to prescribe the surface boundary condition. Additionally,
models below the 0.8 \msun\ boundary are less sensitive to model input parameters such as 
metallicity, the convective mixing length, and age compared to models above that threshold.
The upper boundary to the ``solar-like'' regime at 1.2 \msun\ corresponds, roughly, to the 
mass threshold above which main-sequence stars are believed to develop a convective core.

Definitions outlined in the previous paragraph are the result of knowledge accumulated through 
decades of research on stellar evolution. The importance of many of the above effects were revealed 
over time, some through the study of EBs. It is therefore instructive to begin by placing our 
modern understanding of low-mass stellar evolution theory into perspective with a recapitulation 
of historical progress. Following the historical evolution of stellar evolution theory, a summary 
is provided of where we are today, what is ``state of the art,'' and where EBs are making 
significant contributions to the advancement of stellar evolution models. Finally, ideas about
how EBs may contribute significantly to stellar evolution theory are presented.

Before embarking on the historical perspective, one must caution that the historical review 
presented below attempts to cover the most important advances in the state of low-mass stellar 
evolution theory. As such, individual studies contributing to the the knowledge and inspiration
of those studies cited below may appear under appreciated, as might the effort that goes into 
measuring stellar masses and radii. There is no intent to minimize the meticulous work of theorists
and observers who put tireless hours into advancing physical models and extracting exquisite 
measurements from their data. Without these efforts, theorists would be unable to construct 
advanced computational models and they would be unable to casually remark on the agreement between
observations and theory.

\section{Historical perspective}
\label{sec:history}
The history of testing low-mass stellar evolution theory with EBs begins with the first
observations of YY Geminorum \citep[hereafter YY Gem, also Castor~C;][]{Adams1920} and its
subsequent classification as an EB \citep{vanGent1926,Joy1926}. YY Gem consists of two
M0-type stars that are nearly identical when it comes to their fundamental properties ($M \approx 
0.60$ \msun, $R \approx 0.62$ \rsun, \teff $\approx 3800$ K). Although it would be 
another 20 years before YY Gem provided clues about the inadequacy of stellar model physics, it
will be continually mentioned throughout the historical development of low-mass stellar evolution 
theory.

The first major contribution provided by YY Gem (and EBs) to low-mass stellar evolution 
theory came in the early 1950s. At this point, significant advancements in energy generation
processes via nuclear reactions had been made and the proton--proton ($p$--$p$) chain 
was established as the dominant energy production mechanism in late-type stars 
\citep{Aller1950,Salpeter1952}. However, comparisons of stellar evolution models computed assuming 
energy production via $p$--$p$ chain predicted stellar luminosities far in excess of observed 
luminosities for stars below the Sun in the HR-diagram \citep{Aller1950,Aller1952}. \citet{Stromgren1952} 
identified numerous possible shortcomings in stellar model physics, including the validity 
of the equation of state (EOS) and adopted radiative opacities. Most importantly, Str\"{o}mgren 
noted that the prevailing assumption of radiative equilibrium \citep[e.g.,][]{Eddington1926} 
throughout stellar interiors was likely incorrect and should be abandoned. Instead, it was 
argued that the properties of a surface hydrogen convection zone in late-type stars should 
be explored. 

Following Str\"{o}mgren's suggestion, \citet{Osterbrock1953} computed the first set of stellar
models of late-type stars with an extended outer convective envelope. Osterbrock found 
satisfactory agreement between model calculations with a deep convective envelope and 
observations of YY Gem for models where the hydrogen convection zone comprised the outer
30\% of the star, by radius. The study by Osterbrock, motivated by observational properties
of YY Gem, revealed that stellar models must account for the transport of energy by convection,
a point that seems rather trivial from our modern point of view, but represented a significant
advance in stellar evolution theory, at the time. However, while the inclusion
of an outer convection zone provided agreement between model predictions and the properties
of YY Gem, there was still significant disagreement between models and the properties of 
stars of later spectral type. Most notable were disagreements with the stars in \kr, a 
visual binary consisting to two mid-to-late M-dwarfs (today classified as M2.0 and M4.0).
Although the system is not eclipsing, its role in stellar structure theory is important and 
worth mentioning. Both stars were observed have a luminosity well below the predictions of 
stellar evolution models, much the same as the stars in YY Gem before the introduction of a 
hydrogen convection zone. These remaining errors were initially attributed to missing effects 
of electron conduction as an energy transport mechanism \citep{Osterbrock1953}. 

Carefully ruling out severe observational errors in the determination of stellar properties 
for \kr, \citet{Limber1958a} pointed out that several additional pieces of physics were likely 
necessary to reconcile model predictions with the observations. These physics included partial 
electron degeneracy of the stellar interior and a more rigorous treatment of radiative and
convective temperature gradients for computation of energy transport in stellar interiors.
However, even when all of these effects were included in models, significant disagreements 
with \kr\ remained \citep{Limber1958b}. Instead, Limber extrapolated on the advances provided
by \citet{Osterbrock1953} and allowed the model interiors to be in full convective equilibrium.
Under this assumption, models were able to provide predictions consistent with the properties 
of the stars in \kr\ \citep{Limber1958a,Limber1958b}.

Precise measurement of the fundamental properties of the stars in YY Gem helped initiate the
establishment of
the current paradigm that outer convection zones grow deeper toward later spectral types,
effectively overturning the prevailing hypothesis of radiative equilibrium. The revolution
was so complete that stellar interiors of the latest type stars were determined to be 
fully convective! However, agreement between models and data was foreseen to be short-lived.
Observational data still possessed large error bars that greatly helped to mask the identification
of further disagreement with model predictions \citep{Limber1958b}. At the same time, Limber
warned that measuring stellar radii would lead to more robust comparisons with stellar 
evolution models. To this end, he suggested further advances in theory that may be required, 
including a larger nuclear reaction network to include the burning of lithium and deuterium, 
departures from hydrostatic equilibrium, ``violent atmospheric activity'' now associated with 
magnetic activity, and the influences of rotation and magnetic fields. Yet, despite the obvious
complications that could enter into the picture, Limber urged caution \citep[pg.~368]{Limber1958a}, 
\begin{quote}
{\it We should not attempt to introduce these added complications unless and until the simpler
 models can no longer account for the observations.}
\end{quote}

Building on the framework laid out \citet{Limber1958b} and following the suggestion that 
improvements to simpler models be explored thoroughly, the next significant advancements 
in low-mass stellar evolution theory came with the introduction of the mixing length theory 
(MLT) of convection for stellar models \citep{bv58,Henyey1965} and improved numerical 
schemes for solving the set of stellar structure equations \citep{Henyey1964}. These advances
permitted more detailed stellar models to be constructed. Several groups created sets of 
low-mass stellar evolution models that include the aforementioned advances, as well as more 
detailed stellar atmosphere calculations following a gray $T(\tau)$ construction \citep[e.g.,][]{
Copeland1970,Hoxie1970}. These calculations included treatment of various atomic ionization
states and molecular dissociation on the EOS and used improved opacity data for bound-free 
and free-free absorption.

\begin{figure*}[ht]
	\centering
	\includegraphics[width=1.4\columnwidth]{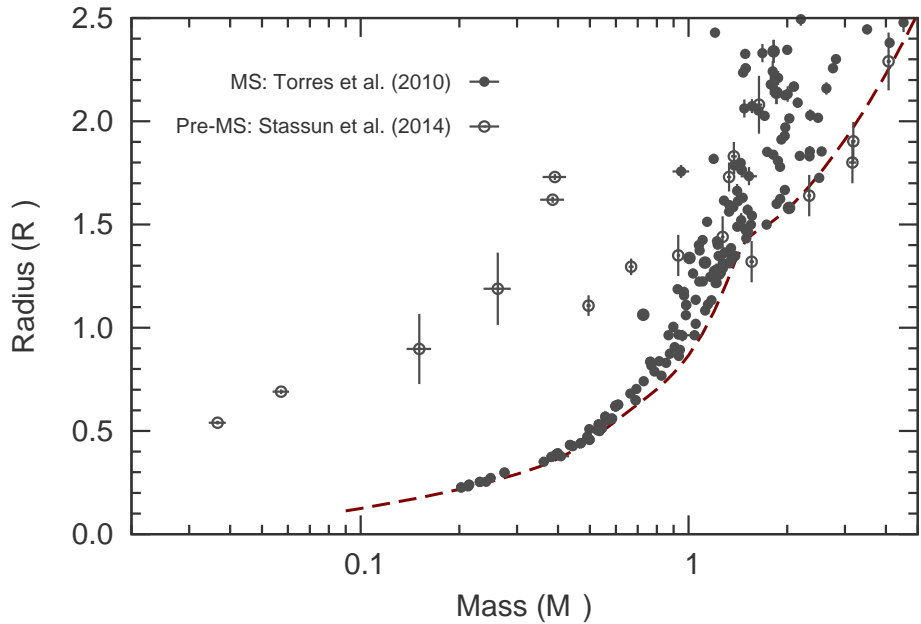}
	\caption{\small Mass--radius diagram for known eclipsing binaries with masses and radii determined to 
	better than 3\% on the MS \citep[filled points;][]{Torres2010,Kraus2011,FC12} and 5\% on the 
	pre-MS \citep[open points;][]{Stassun2014}. The theoretical ZAMS from Dartmouth stellar evolution 
	models is shown as a dashed line.}
	\label{fig:mr_ebs}
	\vspace{-0.2in}
\end{figure*}

Stellar model calculations were found to be in good agreement with the observational mass-luminosity
relationship, but model radii were found to under-predict the radii of stars by up to 40\% 
below about 0.8 \msun\ \citep{Hoxie1970}. Solace was found by noting that observational errors 
were still dominant, masking any disagreements \citep{Hoxie1973}. However, models also 
failed to fit the observationally determined properties of YY Gem within 3$\sigma$ of the
observed radius uncertainties, a fact that was not explicitly mentioned. Instead, YY Gem was used 
to highlight observational uncertainties in the single star effective temperature scale, justifying 
the large uncertainties in the empirical mass-radius relationship. Nevertheless, potential
deficiencies in the mass-radius, and to a lesser extent the mass-luminosity planes motivated
Hoxie to identify the largest sources of uncertainty in low-mass stellar models. Two specific pieces 
of physics were the inclusion of non-ideal effects in the EOS, which would likely affect the
stellar radius predictions, and development of more sophisticated boundary conditions, such
as non-gray model atmospheres, which would have the largest effect on the stellar \teff\ and
luminosity \citep{Hoxie1970,Hoxie1973}.

The next twenty years saw a significant increase in the number of well-characterized EBs
\citep[e.g.,][]{Popper1984}. Unfortunately, only one low-mass EB was added to the list
with YY Gem. That particular system would, however, become just as crucial of a system as
YY Gem for comparisons with stellar evolution theory. CM Draconis was characterized by
\citet{Lacy1977} and was found to contain two very-low-mass stars with masses similar to
those in \kr, meaning they were very likely fully convective throughout their interior
\citep{Limber1958b}, although this was not immediately recognized. The stars in CM Dra 
also appeared to lie well above the theoretical zero-age-main-sequence (ZAMS) for 
Population I objects in a mass-radius diagram, but this was attributed to the fact that 
CM Dra may be a Population II object \citep{Lacy1977}.

The surge of well-characterized EBs during these intervening years were compiled by \citet{Andersen1991}.
In that review, systems were required to have component masses and radii measured with
better than 2\% precision so that they might act as strong tests of stellar evolution
theory. However, there was only one low-mass system in that collection: YY Gem
\citep{Leung1978,Andersen1991}. CM Dra's components had masses 
determined to 4\% and radii to 3\% \citep{Lacy1977} and therefore did not merit inclusion 
in Andersen's compilation.

Low-mass stellar models made their next big leap with the Lyon stellar models \citep{Baraffe1995}. 
What set their models apart from other groups developing models concurrently 
\citep[e.g.,][]{Dorman1989,Burrows1993} was inclusion of an advanced EOS designed 
specifically to accurately model the cool, dense plasma characteristic of low-mass stars 
\citep{scvh95} and adoption of non-gray model atmospheres \citep{Allard1995} to define 
surface boundary conditions for their interior models. The latter feature, in particular, 
allowed their models to accurately treat energy transfer in optically thin regions of the 
stellar atmosphere, where convection and radiation both contribute significantly to the 
overall energy flux \citep{Dorman1989,Burrows1993,Allard1995,Baraffe1995}.
Naturally, one of the first tests of their models was comparing model predictions to the 
properties of YY Gem and CM Dra \citep{Chabrier1995}. Initial results were encouraging, 
showing that their models were able to reproduce the observational properties of both 
systems. However, revisions to the fundamental properties of YY Gem and CM Dra would 
again show models unable to accurately predict stellar properties \citep{Metcalfe1996,
Torres2002}. Nevertheless, the physics advances implemented in the Lyon models 
\citep{CB97,BCAH98} remain state of the art, making them still highly relevant.

\section{State of the art} 
\label{sec:now}
In the years since \citet{Andersen1991}, there was another explosion in the number of  well-characterized 
EBs, especially in the low--mass regime. To synthesize the wealth of observational data, an updated 
compilation of EBs with precisely known properties was published by \citet{Torres2010}. That review 
contains 95 systems, but only five systems possess at least one component mass below 0.8 \msun, and only one system 
contains a star below the fully convective boundary. The lack of low-mass systems was relieved in subsequent 
years thanks largely to photometric surveys searching for exoplanets and variable stars, such as \emph{Kepler} 
\citep{Carter2011,Doyle2011}, MOTESS-GNAT \citep{Kraus2011}, MEarth \citep{Irwin2009,Irwin2011}, All Sky 
Automated Survey \citep[ASAS;][]{Helminiak2011a,Helminiak2011b,Helminiak2012,Helminiak2014}, Super Wide Angle 
Search for Planets \citep[SuperWASP;][]{Triaud2013,Chew2014}, and HATNet \citep{Zhou2014}. Figure \ref{fig:mr_ebs} 
shows a subset of known EBs (see John Southworth's DEBCat\footnote{ http://www.astro.keele.ac.uk/~jkt/debcat/} 
for a complete listing) whose masses and radii are determined with precisions better than 3\% 
and 5\% for main sequence (MS) and pre-MS EBs, respectively.  

\begin{figure}
	\centering
	\includegraphics[width=0.85\columnwidth]{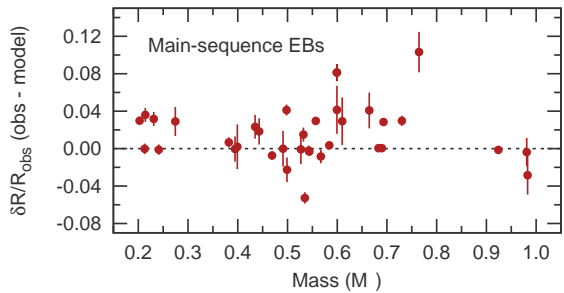}
	\caption{\small Stellar evolution models fail to reproduce the properties of low-mass main-sequence stars 
	in eclipsing binaries \citep[adapted from][]{FC12}.}
	\label{fig:mr_err}
	\vspace{-0.2in}
\end{figure}

Standard stellar evolution models of low-mass stars employ similar physics to the Lyon series: surface
boundary conditions defined by non-gray model atmospheres, modern radiative opacity computations, and 
an EOS that accounts for partial electron degeneracy, Coulomb interactions, and other non-ideal effects.
The rapid influx of high quality data and high precision mass and radius measurements permits a more 
reliable evaluation of stellar evolution model accuracy. Such evaluations indicate that the problems 
identified by \citet{Hoxie1973} still remain, though at a lower level than originally suggested. Canonically,
model radii are quoted to underpredict observed values by between 5\% and 10\% \citep{Ribas2006}. These
values can be decreased to below 5\%, if potential metallicity and age variations are taken into
account \citep{FC12}, as shown in Figure~\ref{fig:mr_err}. However, it is worth noting that the most 
significant outliers are also the most well-characterized systems, including YY Gem (at 0.6 \msun) and CM Dra
(near 0.23 \msun). 

At the same time, it is clear that problems facing stellar evolution theory below 0.8 \msun\ 
are not unique to this mass regime. Williams (this volume), for example, illustrates numerous problems 
that exist with high mass models. Perhaps more worrying, is that studies of EBs with solar-like components 
reveal that models in the solar-mass regime fail to reproduce observations \citep{Popper1997,Clausen2009,vos12}. 
Stellar evolution theorists have previously comforted themselves knowing that, while there are problems 
in various locations of the HR Diagram, models of the Sun and solar-like stars are accurate. If stars in 
EBs are representative of the single star population, these errors can have a profound impact on
interpretation of asteroseismic data, on characterizations of exoplanet host stars, and 
potentially on isochrone fitting to globular and open cluster data.
Given that modern stellar evolution models show systematic disagreements with EB data, what
possible physics may be incomplete or entirely missing from stellar models?

\subsection{Observational evidence}
The first clue
was presented by \citet{MM01}, who noted that low-mass EBs showed similar departures from
standard stellar model isochrones as single active M-dwarfs, a conclusion supported by 
additional comparisons of EBs to the single star population \citep{Morales2008}. This lead
\citet{MM01} to the hypothesize that magnetic activity may be the culprit for the observed
model--observation disagreements. Their hypothesis was supported by the fact that most
EBs had short orbital periods, typically less than three days \citep[due to observational 
biases;][]{Ribas2006}. Nonetheless, stars in these EBs are expected to have rotational 
periods synchronized with the orbital period, meaning these stars are rapidly rotating 
\citep{Zahn1977}. Rapid rotation then drives strong magnetic fields through a hydrodynamic 
dynamo. 

Further evidence in support of the magnetic hypothesis was provided when it was shown that
radius discrepancies between models and observations of low-mass EBs correlate with magnetic
activity in the form of coronal X-ray emission \citep{Lopezm2007}. More active stars, identified
as having a higher ratio of X-ray luminosity to bolometric luminosity, were found to show
larger disagreements with 1 Gyr, solar metallicity Lyon models. In that same study, there appeared
to be no correlation between stellar metallicity and radius deviations for EB systems, a 
trend that was apparent for single stars and was proposed as an alternative hypothesis to 
magnetic activity \citep{Berger2006,Lopezm2007}. This is still some of the strongest evidence 
in support of the magnetic hypothesis.

A second approach to assessing the influence of magnetic activity is to directly compare
model-observation radius disagreements with stellar rotation. Since activity is theorized
to increase with increasing stellar rotational velocity, one expects to see shorter period
binaries display larger radius deviations, while longer period binaries should show better
agreement with model predictions. \citet{Kraus2011} found that short period binaries, with
orbital periods less than 1~day, displayed a significantly higher level of radius inflation
compared to model predictions than those with periods greater than 1 day. This was supported,
in part, by a later study that found a significant break in the level of radius inflation at
1.5 days \citep{FC12}. However, no significant difference was found for EBs with orbital periods
of 1 or 2 days, suggesting the difference may be due to low number statistics, particularly
for systems with orbital periods greater than 3 days. Even accounting for the potential 
influence of rotation on convection, through the use of the Rossby number,\footnote{ Ro = $P_{\rm rot}/\tau_{\rm conv}$,
the ratio of the convective turnover time to the rotational period} no significant correlation
between radius deviations and rotation is observed \citep{FC12}. Curiously, \citet{FC12}
still uncover a correlation with levels of coronal X-ray emission, similar to \citet{Lopezm2007}.
However, their correlation is neither as clear nor as strong as in \citet{Lopezm2007}, and 
it is observed to be driven largely by two data points, including YY Gem.

Presently, there appears to be no clear observational consensus identifying either magnetic
activity or rotation as drivers of the observed radius discrepancies. A few binary systems with 
orbital periods greater than 18 days are now known whose stars show significant radius 
disagreements with stellar models \citep[see, e.g.,][]{FC13}. This does not immediately rule
out magnetic activity and/or rotation as a culprit, but certainly adds a layer of complexity
to problem. These stars in long period EBs, which show radius disagreement, tend to have masses
placing them below the fully convective boundary, perhaps suggesting rotation and magnetic fields 
are driving errors for models of stars with radiative cores and convective outer envelopes
and that models of fully convective stars are plagued by other errors. This evidence has,
in any case, inspired theoretical investigations into the effects of magnetic fields on 
low-mass stellar structure.

\subsection{Theoretical advancements} 
The theoretical basis behind suspecting magnetic fields is that strong magnetic fields can 
suppress the flow of energy across a given surface within a star \citep[e.g.,][]{Thompson1951,GT66}. 
Since stars must maintain a given flux at their surface, commensurate with the energy production 
rate in their core, the inhibition of the outward transfer of flux at any point within the 
star will force the star to readjust to compensate for the suppressed flux. This compensation 
is expected to take the form of an inflated radius and decreased \teff. The same physical 
argument applies to both strong magnetic fields globally suppressing convection \citep[e.g.,][]{MM01} 
or the blocking of radiative flux near the surface by starspots covering some fraction 
of the stellar surface \citep{Spruit1982a}.

\subsubsection{Magnetic suppression of convection}
Stellar evolution models that include parametrized descriptions of the influence of global 
magnetic fields have been constructed by \citet{LS95}, \citet{Ventura1998}, \citet{DAntona2000}, 
\citet{MM01},
and more recently by \citet{FC12b}. Investigation of the effects on low-mass stellar properties
in the context of model disagreements with fundamental stellar properties were carried out
with the models by \citet{MM01} and \citet{FC12b}, the latter of which uses the prescription
derived by \citet{LS95}. Results from both groups suggest that magnetic suppression of convection 
is able to reconcile model radius predictions with observations of low-mass benchmark EBs, 
most notably YY Gem \citep{FC13,MM14} and CM Dra \citep{MM11,FC14}. 

In stars with radiative cores and convective envelopes, such as YY Gem, quantitative 
predictions of surface magnetic field strengths necessary to reconcile model radii appear
consistent with empirical data. Model magnetic field strengths range from a few hundred 
gauss up to a few kilogauss. Direct measurements of magnetic field strengths 
on active, early M-dwarfs confirm these predictions are realistic \citep{Reiners2012a},
as do estimates of magnetic field strengths from indirect indictors, such as chromospheric
Ca {\sc ii} emission and coronal X-ray emission \citep{FC13,MM14}. Interior magnetic 
field strengths predicted by models ($\sim$ 10 -- 100 kG) also appear consistent with 
expectations from more realistic magnetohydrodynamic simulations of stellar magnetic 
fields \citep{Brown2010}. However, models of stars with radiative cores are relatively
insensitive to deep interior magnetic field strengths, as the influence of magnetic fields
in the superadiabatic layer near the stellar surface is of greater consequence \citep{
DAntona2000}. The broad consistency of model magnetic field strengths with empirical data 
suggests that magnetic models may be capturing relevant physics for these types of stars, 
despite simplified magnetic field prescriptions.

The situation for models of fully convective stars, typified by CM Dra, is more complex.
Surface magnetic field strengths predicted by models span a similar order of magnitude, 
anywhere between about 0.5~kG to 5.0~kG \citep{MM11,FC14}. These values are reasonable
when compared to average surface magnetic field strengths of mid-to-late field M-dwarfs
\citep[$\sim$ 3 kG; e.g.,][]{Reiners2007}.
Debate about whether magnetic fields are actively inflating fully convective stars largely 
focuses on predicted interior magnetic field strengths \citep{Chabrier2007,MM11,FC14}.
Unlike stars with radiative cores, models of fully convective stars are less sensitive to
the strength of the magnetic field in the near-surface layers and appear to be relatively
more influenced by interior magnetic field strengths \citep{MM01,MM11,FC14}. Interior magnetic
field strengths typically need to be in excess of 1.0~MG. Interested readers are encouraged 
to consult the aforementioned references for a full account of the arguments for and against
the appearance and maintenance of megagauss magnetic fields in fully convective stars. 
It is likely that stronger constraints from magnetohydrodynamical simulations are needed 
to further the debate. Nevertheless, there is some concern over the magnitude of interior 
magnetic field strengths and, thus, global magnetic fields as a solution for the mass-radius
disagreements between models and observations.

\subsubsection{Magnetic activity -- starspots}
Alternatively, it may not be global magnetic fields inhibiting convective flows that suppresses
flux and forces stars to inflate, but intense local concentrations of magnetic fields producing
starspots on the stellar photosphere. Magnetically active stars show modulations in their 
lightcurves due to the presence of dark spots.\footnote{ There is the distinct possibility that 
lightcurve modulations are caused by bright spots, but it has been argued that this is unlikely 
\citep{Berdyugina2005}.} Dark spots reduce the radiative output from the stellar surface by 
trapping excess energy at their base \citep{Spruit1982b}. Depending on the lifetime of a given 
spot and the efficiency at which the trapped energy can be redistributed, a star may grow larger 
and cooler in response to the presence of spots \citep{Spruit1982a}. 

It was shown that, in the case of low-mass EBs, effects due to spots may be more significant in
driving structural changes in a stellar model than effects from global inhibition of convection 
by magnetic fields \citep{Chabrier2007}. This is of particular consequence for models of fully
convective stars, like CM Dra, where global inhibition of convection may not be sufficient, at
least with realistic magnetic field strengths \citep{Chabrier2007,Morales2010}. However, starspots
may have an additional influence on studies of EBs. The presence of starspots can bias radius 
measurements from EB lightcurves (largely through biasing of the measured radius sum) toward 
larger radii by up to 6\%, depending on the properties of spots \citep{Morales2010,Windmiller2010}. 
For systems like YY Gem and CM Dra, \citet{Morales2010} estimated that the stellar radii may be 
overestimated by about 3\%. It is not possible to necessarily attribute all of the observed errors 
to this bias, but assuming these stars have smaller radii assists theoretical models of radius 
inflation through magneto-convection or starspot flux suppression by lowering the required amount 
of radius inflation \citep{Morales2010,MM11}. 

Critical to the idea that spots bias radius measurements is that spots are preferentially located
near the rotational poles and cover a significant fraction of the stellar surface (between 40 -- 
60\%). Suppression of flux by spots at the surface is not sensitive to distribution, but is highly
dependent on surface coverage and spot temperature contrasts. Presently, there is no strong empirical
evidence for polar spots on either rapidly rotating low-mass stars or fully convective stars 
\citep[see Section 4.3 in][for an in-depth discussion]{FC14}. In fact, there is evidence that spots 
must be randomly distributed across the surface of fully convective stars to produce the diversity 
of lightcurve morphologies for stars in the young open cluster NGC 2516 
\citep{Jackson2013}. This leads to the possibility that fully convective stars have random 
distributions, but large filling factors (total surface coverage). Observations of FeH molecular band
features in spectra M-dwarfs reveal that M-dwarfs are probably covered nearly everywhere by 1 kG
magnetic fields, with small patches of intense 5 kG -- 8 kG magnetic fields, averaging to about 3 kG 
over the surface \citep{Shulyak2011}, supporting this idea. However, this assumes a one-to-one correlation 
between the presence of 1 kG magnetic fields and the appearance of spots. Investigations of solar 
magnetic fields reveal that 1 kG magnetic fields may not be sufficient to inhibit
convective flows \citep{Mathew2004}. Convective flows in low-mass stars likely react differently
to the presence of magnetic fields than convective flows in the Sun, but it raises the question of 
whether starspot properties required by stellar models are realistic.

\section{Exploring alternative solutions}
While magnetic fields and activity are able to provide a solution to the mass--radius 
problem with low--mass stars, questions remain about the reality of stellar evolution
model predictions. These questions will undoubtedly be answered by future observations. 
However, perhaps an equally valid epistemological approach is to explore alternative 
theoretical solutions with the aim of removing the need to invoke magnetic fields, or 
with the aim of ruling out the other alternatives and thus bolstering arguments in 
favor of the magnetic hypothesis. Recalling discussions in Section~\ref{sec:history},
this is the approach advocated by \citet{Limber1958a}.

\subsection{Convection}
\label{subsec:conv}
One alternative to the magnetic hypothesis is that convection in low-mass stars is 
significantly less efficient than it is in more solar-like stars. This argument has
been explored previously, but investigations have largely focused on individual systems. 
Suggestions for why convective properties may be different among individual EBs that 
show radius disagreements with models include magnetic fields \citep{Cox1981}, 
rotation \citep[Coriolis force acting on convective flows;][]{Chabrier2007}, and 
intrinsic differences due explicitly to stellar properties \citep[i.e., mass, metallicity, 
\teff;][]{Lastennet2003}. While it is important to demonstrate that manipulating 
convective properties---here largely the convective MLT parameter $\amlt$---provides 
relief to noted model--observation disagreements, it makes identifying physical 
explanations for those manipulations difficult. Instead, statistical properties 
from a sample of EBs provides an opportunity to reveal meaningful trends that betray
physics associated with changes in stellar convective properties.

Such a study was recently performed for solar-like EBs \citep{Fernandes2012}, where
$\amlt$ was manipulated to provide the best agreement between models and EBs. Results
show that $\amlt$ does not appear to be mass dependent, as stars of equal mass from 
different EB systems require different values for $\amlt$. Instead, \citet{Fernandes2012} 
demonstrate that $\amlt$ correlates with \vsini, where faster rotating stars 
require lower $\amlt$ values. While the results are tantalizing, comparison of models
against solar-like EBs introduces errors due to unknown stellar ages and compositions
(both helium abundance $Y$ and metal abundance $Z$), which must be simultaneously fit. 
Although stars in a given EB must lie along the same isochrone, which isochrone is 
assumed correct can decidedly alter modeling conclusions. Extending this type of study 
to low-mass EBs would, however, help relieve modeling uncertainties associated with 
unknown stellar ages and compositions, as low-mass stellar models are relatively less 
sensitive to assumptions about these parameters. Having at least some constraint on 
stellar metallicity can drastically improve the quality of model comparisons, so 
observers are anyway encouraged to put forth effort to measure stellar metal content. 

It is quite remarkable that, through a systematic comparison of the low-mass EB population
to stellar models, it may be possible to extract information regarding the hydrodynamic 
properties of low-mass stars as a function of a range of variables, such as mass, [$M$/H], 
\teff, and \vsini\ in a similar manner as is done with asteroseismology \citep{Bonaca2012}.
This knowledge would lead to significant improvements in the treatment of convection
in stellar evolution models and would represent yet another facet of stellar physics
advanced by studies of EBs.

\subsection{Heavy element composition}
Stellar metallicity plays only a minor role in modern discussions of the low-mass
EB mass-radius problem. However, it is not clear, at least to the author, that 
metallicity effects are not partially responsible for the observed disagreements.
Although \citet{Lopezm2007} failed to identify a correlation between metallicity 
and radius inflation among low-mass stars in binaries, the lack of a trend was
largely driven by the points associated with CM Dra. Revisions to the fundamental 
properties of CM Dra \citep{Morales2009a} and recent estimates of its metallicity
\citep{Terrien2012,Kuznetsov2012} have led to a shift in the location of CM Dra 
in the metallicty--radius inflation diagram, revealing circumstantial
evidence that stellar metallicity may be related to observed radius errors
among fully convective stars \citep{FC13b}. 

However, correlations with metallicity among fully convective stars in EBs 
does not follow the same pattern as was observed for single stars \citep{Berger2006}. 
Instead of more metal rich stars showing larger radius inflation, radius errors 
among fully convective stars in EBs appear to negatively correlated with stellar 
metallicity, with more metal poor stars displaying greater radius errors. This
may point toward errors in the model EOS or opacities. However, the sample
is small, with only six stars in four systems, and is largely driven
by the location of CM Dra in the diagram. The notion is bolstered slightly
by the fact that it does not matter what metallicity is adopted for CM Dra, 
it still falls along the observed negative correlation. It should be noted 
that a recent investigation by \citet{Zhou2014} claims to find no trend of
radius inflation with metallicity among fully convective stars in EBs. However,
their most precisely measured star, and their most metal-poor ([$M$/H] $= -0.6$ 
dex), falls along the relation suggested by \citet{FC13b}. Confirmation of the
properties of other stars in their sample is required. As the number of 
fully convective stars in EBs with metallicity estimates is increased, it will
become clear whether hints of the trend are spurious or not.

\subsection{Helium abundance}
An interesting possibility is to use low-mass EBs to investigate stellar helium 
abundances \citep{Limber1958b}.
Helium abundances in low-mass stars cannot be measured directly, and therefore 
specification of $Y$ in stellar models relies on prior assumptions regarding the
relationship between $Y$ and $Z$. The significant role that helium plays in
governing the structure and evolution of low-mass stars, coupled with our ignorance 
of $Y$ in low-mass stars, makes it a noteworthy suspect in our attempts to solve the
mass-radius problem \citep{Valcarce2013}. Although observational confirmation of
model predicted $Y$ values cannot be obtained, consistency with other measures of
local helium abundances in the solar neighborhood can be sought. 

As with studies of adjusting $\amlt$, previous investigations largely focused on 
individual systems, such as CM Dra \citep{Paczynski1984, Metcalfe1996} and UV Psc 
\citep{Lastennet2003}. Populations of solar-like EBs were used to probe the helium 
enrichment as a function of metallicity in the solar neighborhood \citep{Ribas2000,
Fernandes2012}. It is usually assumed that helium abundance is linearly proportional 
to metallicity, such thats
\begin{equation}
Y(Z) = Y_{\rm P} + \frac{\Delta Y}{\Delta Z}\cdot Z,
\end{equation}
where $Y_{\rm P}$ is the primordial helium abundance. Solar-like EBs suggest that
the slope of the relation is  $\Delta Y / \Delta Z = 2\pm1$ \citep{Ribas2000,Fernandes2012}, 
consistent with 
estimates from single K-dwarfs in the solar neighborhood \citep[e.g.,][]{Casagrande2007}.
However, the inferred primordial helium abundance from these studies is well below 
the primordial abundance estimated from Big Bang Nucleosynthesis \citep[$Y_P \approx 0.225$
compared to $Y_{P,\, \rm BBN} \approx 0.249$;][]{Peimbert2007}.

Inferring helium abundances involves simultaneously fitting multiple stellar model 
parameters (i.e., $\amlt$, $Y$, $Z$, age) to provide the best possible agreement 
between model predictions and observed properties of stars in EBs. It was mentioned 
in Section~\ref{subsec:conv} that these parameters have a strong impact on stellar 
model calculations of solar-like stars, introducing degeneracies in the optimization 
problem. Presently, very few EBs have metallicity estimates, making it difficult to
add observational priors in the optimization scheme. Metallicity estimates would
help tremendously, particularly for solar-like stars. At the same time, with the 
recent increase in the number of well-characterized low-mass EBs, it is now
possible to use low-mass EBs to constrain $Y(Z)$. Models are considerably less sensitive 
to fit parameters, particularly age, as compared to solar-like stars. Metallicities 
are still a concern, but methods of estimating bulk metallicities for low-mass stars 
are showing promise \citep{RojasAyala2012,Mann2013a}. Assuming how helium abundances 
scales with metallicity is a crucial in stellar evolution modeling, and studies
of low-mass EBs provide the best chance to reveal this relation.

\subsection{Probing internal structure}
Perhaps the ultimate test of stellar evolution theory that EBs can provide is a 
direct inference of a star's internal density structure through the measurement 
of apsidal motion in EBs with eccentric orbits. This is very fitting for this 
occasion, where we gather to remember the contributions to binary star science 
by Zden\v{e}k Kopal. 
Observing the precision of an EB orbit's periastron with sufficient accuracy and
precision yields an average interior structure constant, $\overline{k_2}$ of the two 
component stars. For any individual star, $k_2$ quantifies the object's 
central mass concentration \citep{Kopal1978}. Objects that are very centrally 
condensed, with tenuous outer layers and dense cores compared to the average density 
(i.e., high mass stars) are described by lower $k_2$ values than objects that have 
a more even distribution of mass (i.e., M-dwarfs).

The fact that observations only reveal the component averaged $k_2$ value means
that equal mass binaries are preferable. Fortunately, low-mass binaries are 
more likely to be of equal mass (Bate, this volume). However, low-mass stars have
a lower fraction of binarity and typically form with smaller semi-major axes (see 
Bate, this volume), meaning they will tend to rapidly circularize from tidal 
interactions \citep{Zahn1977}. All considered, the properties of low-mass binaries
makes the chances of finding suitable systems for deriving $k_2$ quite low. 

Despite this, if apsidal motion in a low-mass binary is accurately measured, we will 
learn a great deal about the accuracy of stellar models. Primarily, we will learn 
whether models predict accurate density profiles for the deep interiors of low-mass 
stars. The apsidal motion constant is largely determined by the deep interior structure,
with smaller influences from the near surface layers, where the density in individual 
layers is much less than the average density of the star. If models accurately predict
$k_2$ for stars where models do not predict accurate fundamental properties (radius, \teff),
then the problems can be isolated to the near surface layers. In contrast, if the 
deeper layers in models are found to be inadequately described, this points towards
a different set of physics, such as opacities and the EOS. At least one low-mass 
binary gives us this opportunity, KOI-126 \citep{Carter2011}, where the apsidal motion
constants may be determined for two fully convective stars with a precision of 1\%.

\section{Conclusion}

Our current understanding of modeling errors are very much like the early view 
on M-dwarf stars \citep[commenting on \kr]{Russell1917},
\begin{quote}
{\it It is obvious that these stars exhibit every characteristic which might 
be supposed typical of bodies at the very end of their evolutionary history,
and on the verge of extinction.}
\end{quote}
At the time, this view of M-dwarfs was well justified, but nonetheless it
now appears rather strange, as M-dwarfs are thought to live for hundreds of 
billions of years, meaning they are actually at the very beginning of their
evolutionary history. Indeed, it is very easy to grow confident that observed 
errors between models and stars in EBs are themselves becoming extinct. However, 
as history of comparing models to theory has continually demonstrated, more 
often than not, errors persist even with the most sophisticated models. It 
is entirely likely that, as M-dwarfs are actually at the very beginning
of their lives, we are only at the beginning of our journey of reconciling 
errors between stellar evolution theory and observations. It is clear, though,
that well-characterized EBs, like YY Gem, will continue to play critical roles
in advancing stellar evolution theory.

\section*{\small Acknowledgments} 
{\small \noindent Thank you to the science and local organizing committees of 
the Kopal 2014 conference for a successful conference in Litomy\v{s}l 
and for the invitation to present this short review. These ideas have 
benefitted from numerous discussions with B.~Chaboyer, G.~Torres, T.~Boyajian,
R.~Jeffries, A.~Dotter, B.~Gustafsson, and O.~Kochukhov, but any mistakes 
are my own.}

\end{document}